\documentclass[epj,twocolumn]{webofc}
\usepackage[varg]{txfonts}   
\woctitle{NSRT15}
\begin{document}
\title{SOFT PHOTON YIELD IN NUCLEAR INTERACTIONS}
%
%
\subtitle{On behalf of the SVD-2 Collaboration}
\author{E.~Kokoulina\inst{1}\fnsep\thanks{\email{kokoulin@sunse.jinr.ru}} 
}

\institute{JINR, Dubna, Russia and
GSTU, Gomel, Belarus
          }

\abstract{
 First results of study of a soft photon yield at Nuclotron (LHEP, JINR) in nucleus-nucleus collisions at 3.5 GeV per nucleon are presented. These photons are registered by an BGO electromagnetic calorimeter built by SVD-2 Collaboration.  The obtained spectra confirm the excessive yield in the energy region less than 50 MeV in comparison with theoretical estimations and agree with previous experiments at high-energy interactions.
}
\maketitle
\section{Introduction}
\label{intro}
Physicists consider that at interactions of nuclei with proton 
and deuteron beams are formed cold nuclear matter \cite{CNM}. 
These researches permit to compare properties of hot quark-gluon medium 
formed in collisions of relativistic heavy ions and cold nuclear matter producing  
in $pp$ or $p(d)$A interactions. The SVD-2 Collaboration carries out 
studies of $pp$, $p$A and AA interactions. These experiments are fulfilled 
at U-70 in IHEP, Protvino  with 50 GeV-proton beams \cite{Therm} and 
at Nuclotron (JINR) with 3.5 GeV/nucleon nucleus beams. 

The SVD-2 Collaboration investigates the unique region of 
high multiplicity in which a number of phenomena of collective 
phenomena of secondary particles is predicted. It has been received 
the following results \cite{Therm,Fluct}:

$\ast $ the topological cross sections in the region up to 
charged multiplicity
 $N_{ch}$ = 24 that had permitted to advance in them 
 down on three orders of magnitude;
 
 $\ast $ the average multiplicity and variance of number of charged particles;
 
 $\ast $ distributions on the number of neutral pions at fixed total pion multiplicity, 
 $N_{tot} = N_{ch} + N_0$ ($N_0$ -- number of $\pi ^0$-mesons);
 
 $\ast $ the rapid growth of the scaled variance, $\omega =D/\overline N_0$, 
 with the increasing of total pion multiplicity 
 ($D$ -- a variance of the neutral meson number at the fixed total multiplicity,
 $\overline N_0$ -- their mean multiplicity).
 
 The growth of the experimental value $\omega $ 
gets 7 standard deviations in respect of Monte Carlo predictions. This result is 
 one of the evidences of Bose-Einstein condensate (BEC) formation \cite{Gor}.
  
The theoretical description of this collective phenomenon 
has been developed by Begun and Gorenstein~
\cite{Gor} at the specific conditions of the SVD-2 
experiment at U-70 in $pp$ interactions. They have estimated 
the total pion multiplicity at which BEC can start to form.
The German physicist S. Barshay predicts \cite{Bars} that the pion condensation may be accompanied by 
an increased yield  of soft photons (SP) with energy smaller than 50 MeV. 
The anomalous SP have being studied experimentally during more than 30 years  \cite{Chlia,WA83,HELIOS,Perep1}. 
There are some theoretical models worked out for  an explanation of the SP yield 
\cite{Van,Wong,GDM}. Unfortunately, an incompleteness of data does 
not permit disclosing of the physical essence of this phenomenon completely. 

To understand nature of the SP formation more comprehensive 
and in particular to test a connection between 
their excess yield and the BEC formation, a SP electromagnetic calorimeter (SPEC) 
has been manufactured and tested by SVD Collaboration at U-70 \cite{SPEC} accelerator. 
This calorimeter is a stand-alone device and it differs from many similar ones 
by its extremely low energy threshold of gamma-quantum 
registration -- of order of 2 MeV. The SPEC technique permits to execute 
the unique research program of $pp$, $p$A and AA interactions  
with registration of SP. 

The report is organised in the following way. The few previous
SP experiments are reviewed in section 2. In section 3 
the description and technical characteristics of electromagnetic  calorimeter
manufactured by SVD Collaboration are given. The first preliminary spectra of SP obtained with the
deuterium  and lithium beams on a carbon target at Nuclotron are also presented in this section.

\section{Review of some experiments recording soft  photons}
Experimental and theoretical studies of direct photon production in hadron 
and nuclear collisions essentially expand our 
knowledge about multi-particle production mechanisms. 
These photons are useful probes for an investigation of nuclear matter at all stages of the interaction.			
SP play a particular role in these studies. 
Until now we do not have total explanation for the experimentally observed excess of SP yield. 
These photons have low transverse momenta $p_{T}$  <  0.1 GeV/c 
and Feynman variable |x| < 0.01. 
In this domain their yield exceeds the theoretical estimates 
by 3 $\div $ 8 times \cite{Chlia,WA83,HELIOS,Perep1}. 

This anomalous phenomenon has been discovered at the end of 1970s with the 
Big Europe Bubble Chamber at the SPS accelerator, in CERN, 
in the experiment with 70 GeV/c 
$K^+$-meson and antiproton beams \cite{Chlia}.  The SP yield 
had exceeded the theoretical predictions by 4.5 $\pm $ 0.9. 
The following electronic experiments such as \cite{WA83, HELIOS, Perep1} 
have confirmed an anomalous behaviour of SP.  

WA83 Collaboration  studied the direct SP yield at 
OMEGA spectrometer in $\pi ^-$ + $p$ interactions 
at hydrogen target with 280 GeV/$c$  $\pi^-$-mesons. 
Excess yield of SP turned out  to be equal to 7.9~$\pm $~1.4 \cite{WA83}. 
Last experimental study of SP had been carried out at the LEP accelerator 
with DELPHI setup in CERN \cite{Perep1}. 

Two kinds of processes were  investigated in this
experiment:
$e^+ + e^-\to Z^0 \to $ jet~+~$\gamma $ and 
$e^+$ + $e^- \to \mu ^+ + \mu ^- $. In processes with formation of hadron jets 
the DELPHI Collaboration had revealed surpluss of SP yield over of Monte Carlo
estimations at the level 4.0 $\pm $ 0.3 $\pm $ 1.0 times. For the first time the SP 
yield at maximum number of neutral pions 7-8 had amounted to about 17-fold exceeding 
\cite{Perep1} in comparison with bremsstrahlung of charged particles.
On the contrary, in the lepton process without formation of hadron jets
the yield of SP turned out to agree well with theoretical
predictions.

The theoretical models try to explain anomalous yield of SP.
The SVD-2 Collaboration has developed a gluon dominance 
model \cite{GDM} explained  an excessive SP yield by the production
of soft gluons in quark-gluon system. These gluons 
do not have enough energy to fragment
into hadrons, so they are scattered on the valency quarks 
of secondary particles and form SP \cite{GDM}. 
This model gives two-three-fold exceeding over common 
accepted area of strong interactions in accordance 
with estimations of the region of SP emission \cite{GDM}.

\begin{figure}
    \centering
\includegraphics[width=0.5\textwidth]{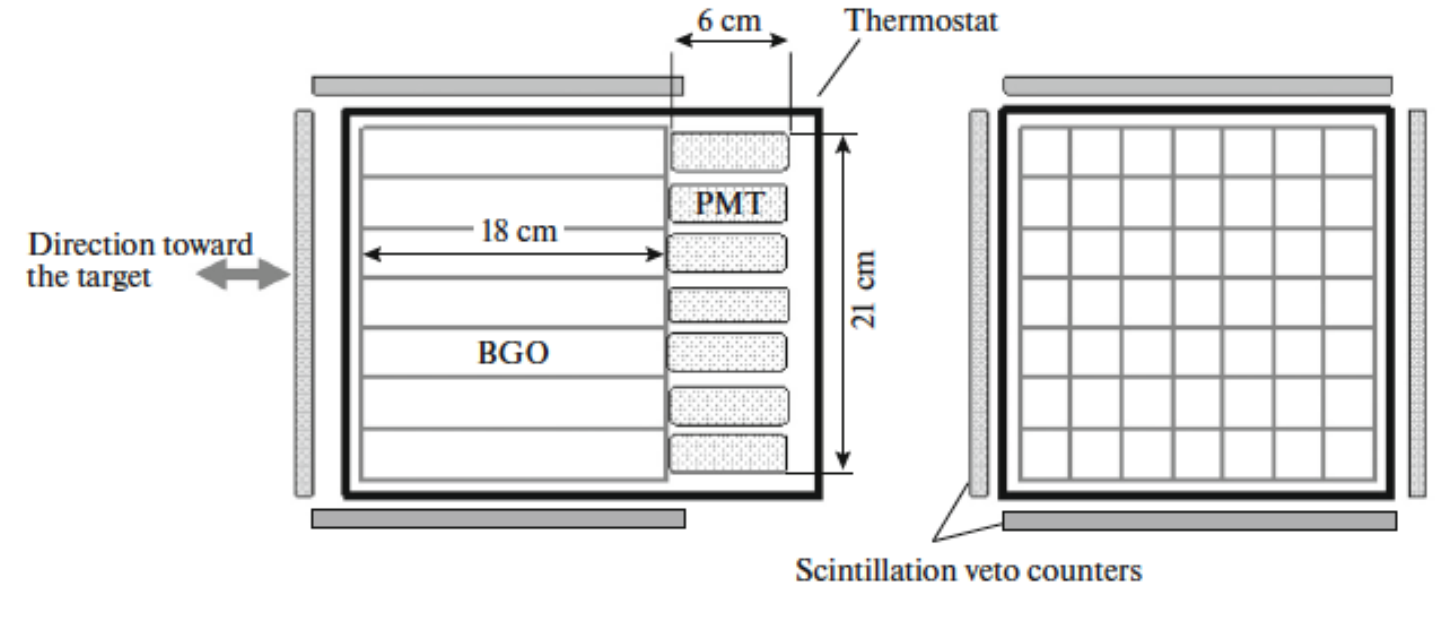}
\includegraphics[width=0.25\textwidth]{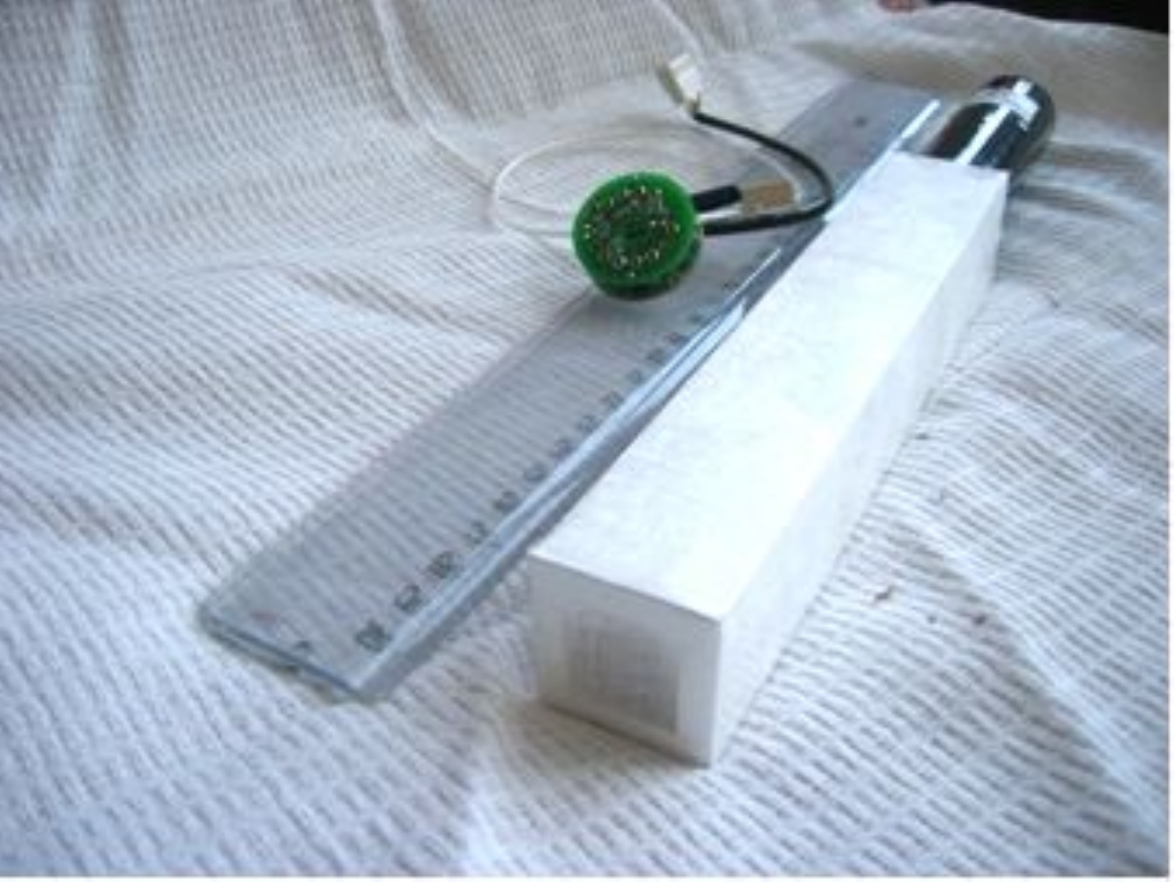}
\caption{Top panel: scheme of SPEC.  Bottom panel: 
BGO crystal with PMT.}
\label{fig-1}
\end{figure}

\section{Technical characteristics of SPEC. Experimental spectra of soft photons}
SPEC has been manufactured 
on the base of the BGO scintillators (bismuth ortogermanate) \cite{SPEC}.
The BGO crystals  
have a small radiation length X$_0$ = 1.12 cm 
(for comparison, crystals NaI(Tl) have X$_0$ = 2.59 cm), 
that permits reducing  considerably the volume of the device. 
Moreover, this scintillator has small sensibility to neutrons. 
It is important at the measurement of a gamma-radiation. 
At manufacturing of such calorimeter 
the problems of uniform distributions of activator 
in the crystal volume do not appear. While many of 
inorganic scintillators have the long-term of emission, 
BGO crystals shows relative small afterglow. 

The scheme of  SPEC is shown in Fig.~\ref{fig-1} (from a top).
It is a square matrix composed 
of 49 (7$\times $7) counters.  
The counter  with 
a demultiplier and a preliminary amplifier
are placed directly on the panel PMT 
as shown  in Fig.~\ref{fig-1} (from below) \cite{SPEC}. 
Every crystal has the parallelepiped 
form, 30$\times $30$\times $180 mm$^3$. 
180 mm correspond to 16 radiation lengths.
Its front side is covered by the high-reflective film VM2000. 
The lateral facets of crystals are wrapping up of Tyvic  
for increasing of light gathering (the thickness 120~$\mu $m).
The PMT 9106SB are used (ET Enterprises). They have 8 dynodes 
and high quantum efficiency in the green part of spectrum. 
The photocathode diameter is equal to 25 mm. 
The tube has the permalloy magnetic protection. 
PMT is glued to the crystal by the optic EPO-TEK 301 glue. 

The box with counters is surrounded 
by the scintillator detectors of a guard veto-system.  
SPEC is placed inside of the thermostat (the top panel of Fig.~\ref{fig-1}). The thermo stabilisation 
is realised by cooling system Huber 006B. The 
temperature of liquid in thermostat can change in the diapason from -20$^\circ $ 
up to +40$^\circ $ C. After few trials the temperature has been chosen +18$^\circ $ C.

\begin{figure}
    \centering
\includegraphics[width=0.5\textwidth]{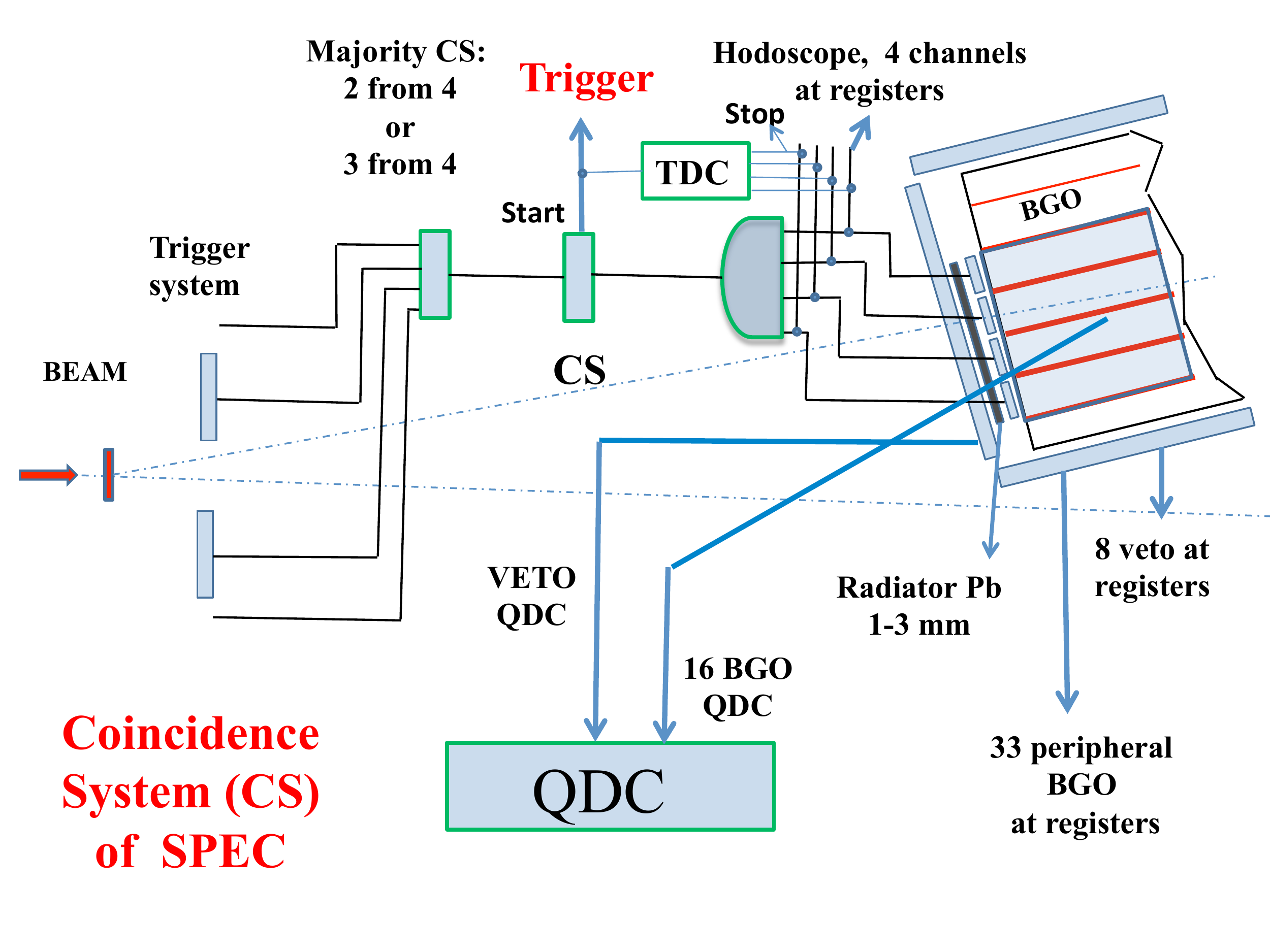}
\includegraphics[width=0.5\textwidth]{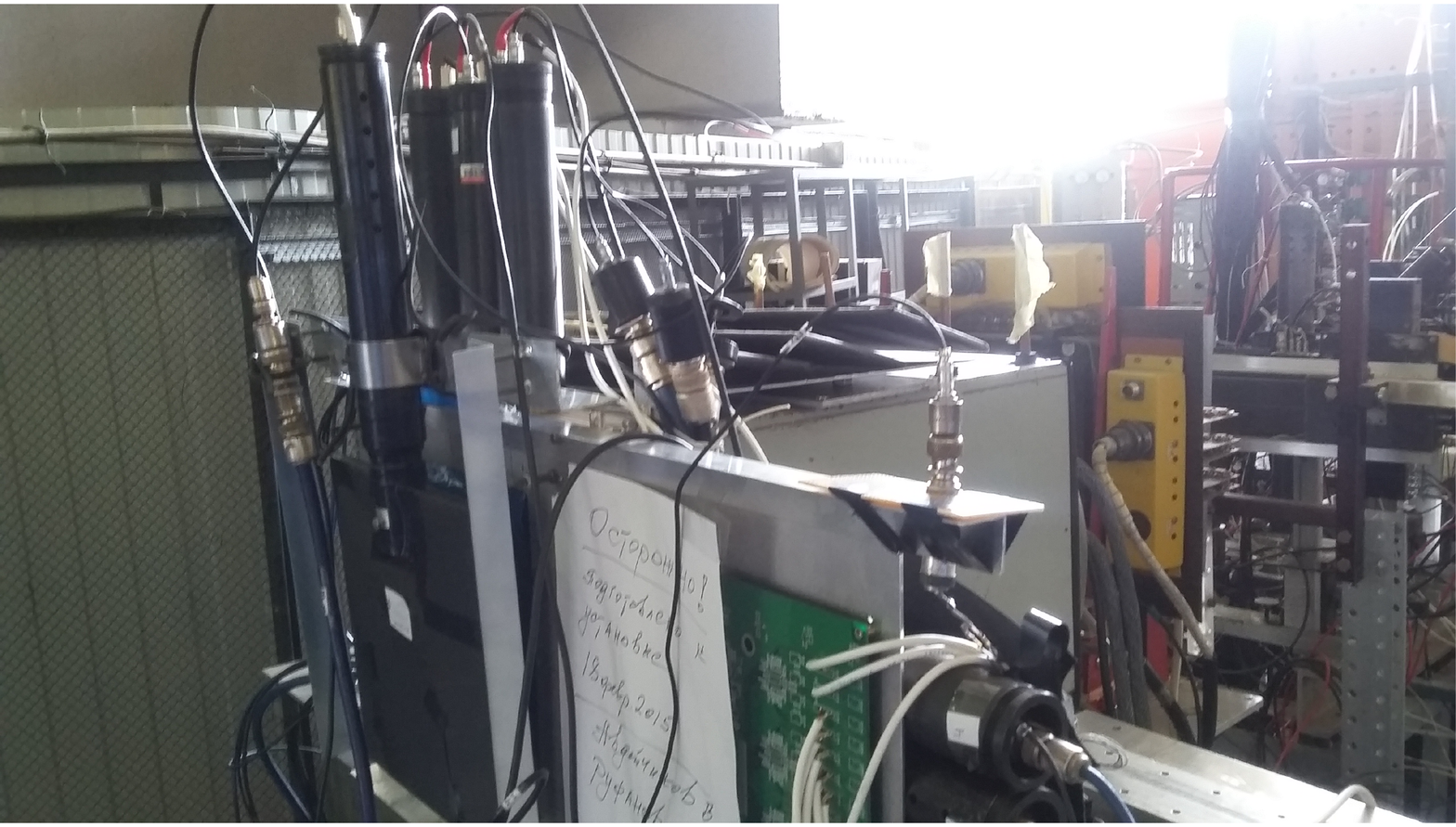}
\caption{Top panel: the reductive working  scheme of SPEC at the Nuclotron. 
Bottom panel: view of SPEC with veto-detectors at NIS-GIBS setup.
}
\label{fig-2}
\end{figure}

The plastic veto-detector of charged particles (23$\times $23$\times $1 cm$^3$)
is placed before the crystals. Behind it an assembly of 4 plastics of a pre-shower 
(18$\times $4.5$\times $1 cm$^3$) is shown in Fig.~\ref{fig-2} (from top). A lead 2 mm-convertor  
is put between the front-veto and plastics. In this figure, the target and
counters of a trigger system are shown also.  The trigger is
produced at the signal from any 2 of 4 pre-shower counters. 
There are two large veto-counters in front of the target. 
They are necessary to forbid a response
from the beam halo. SPEC with veto-detectors
has been laid at Nuclotron hall near the NIS-GIBS setup (Fig.~\ref{fig-2}, bottom panel).

The maximum of the signal-noise ratio is provided by input 
capacitance minimisation. It is determined by the dynode-anode gap 
and capacitance of assembly that  amounts to $\sim $ 6 pF.  
The noise in the spectrometric channels does not exceed 
100 keV that permits for the first time measuring of SP spectrum in 
the range 0.6 $\div $ 600 MeV. The dynamic diapason of 
signals amounts to more than 66 dB.
Time-stamp is given by the 4.5$\times $4.5$\times $0.1 cm$^3$ beam counter 
(not shown). It is also placed in front of the target. 

\begin{figure}
    \centering
\includegraphics[width=0.45\textwidth]{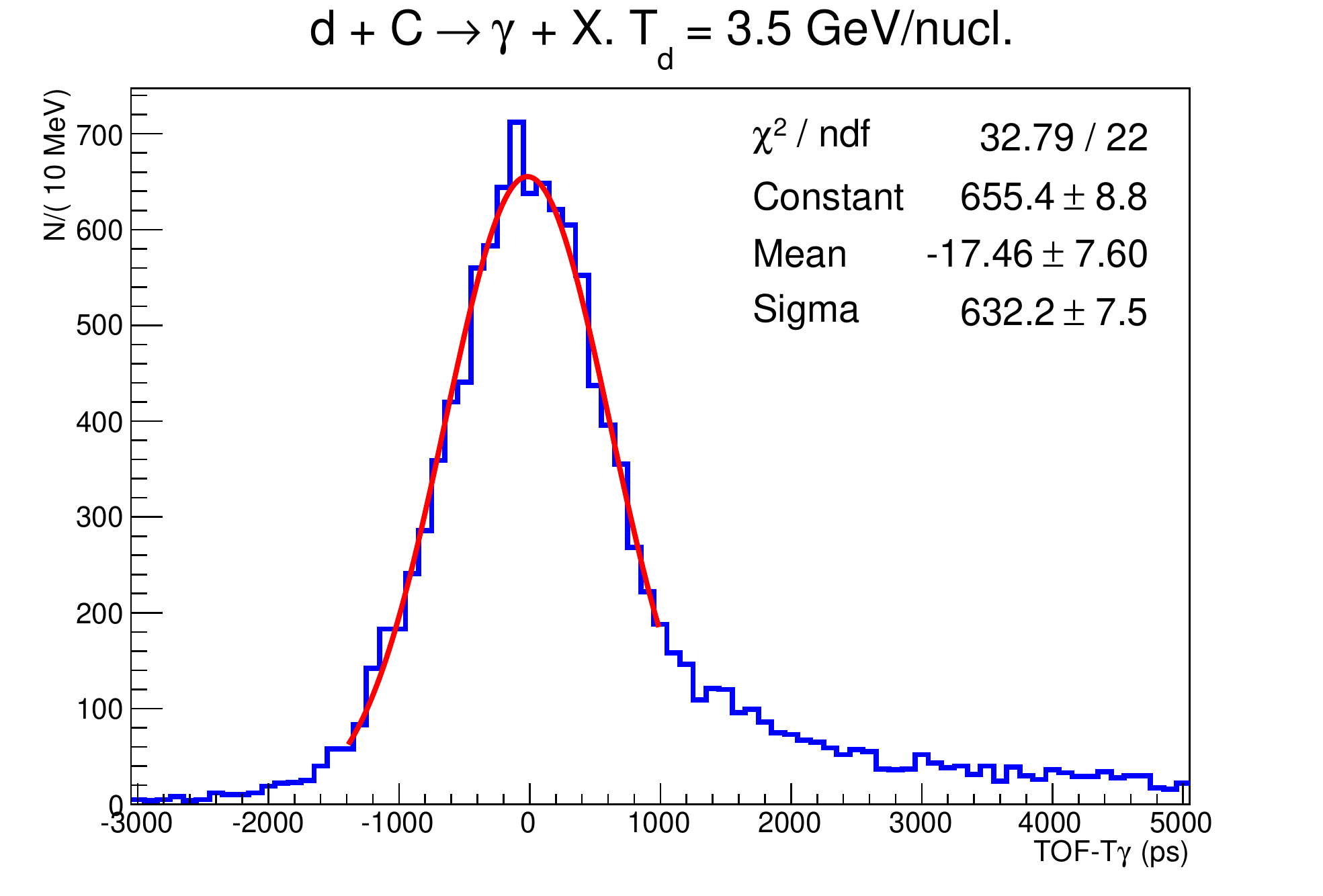}
\includegraphics[width=0.45\textwidth]{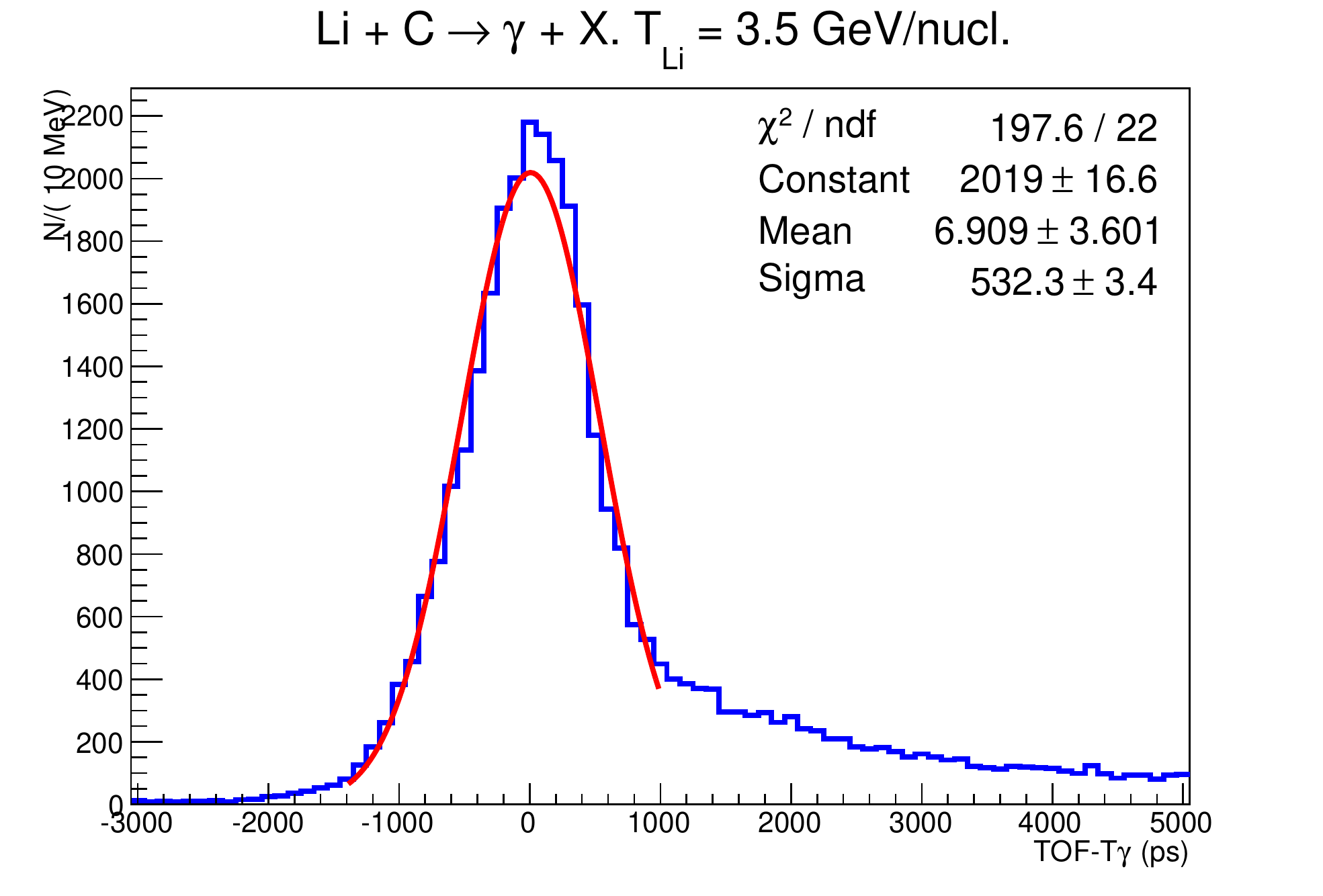}
\caption{Top panel: time resolution in pre-shower for d+C
interactions.
Bottom panel: same as above for Li+C
interactions.}
\label{fig-3}
\end{figure}

\begin{figure}
    \centering
\includegraphics[width=0.45\textwidth]{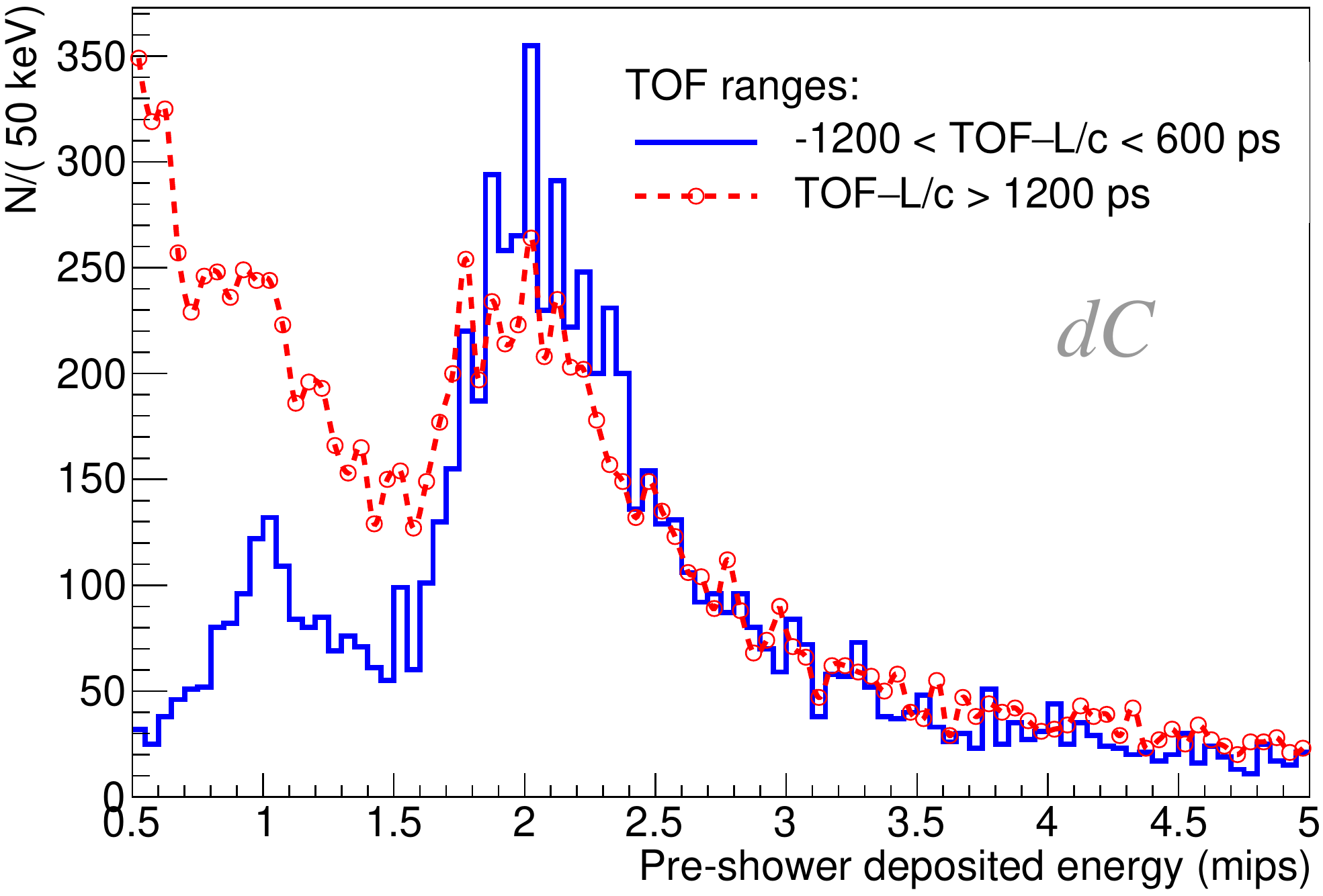}
\includegraphics[width=0.45\textwidth]{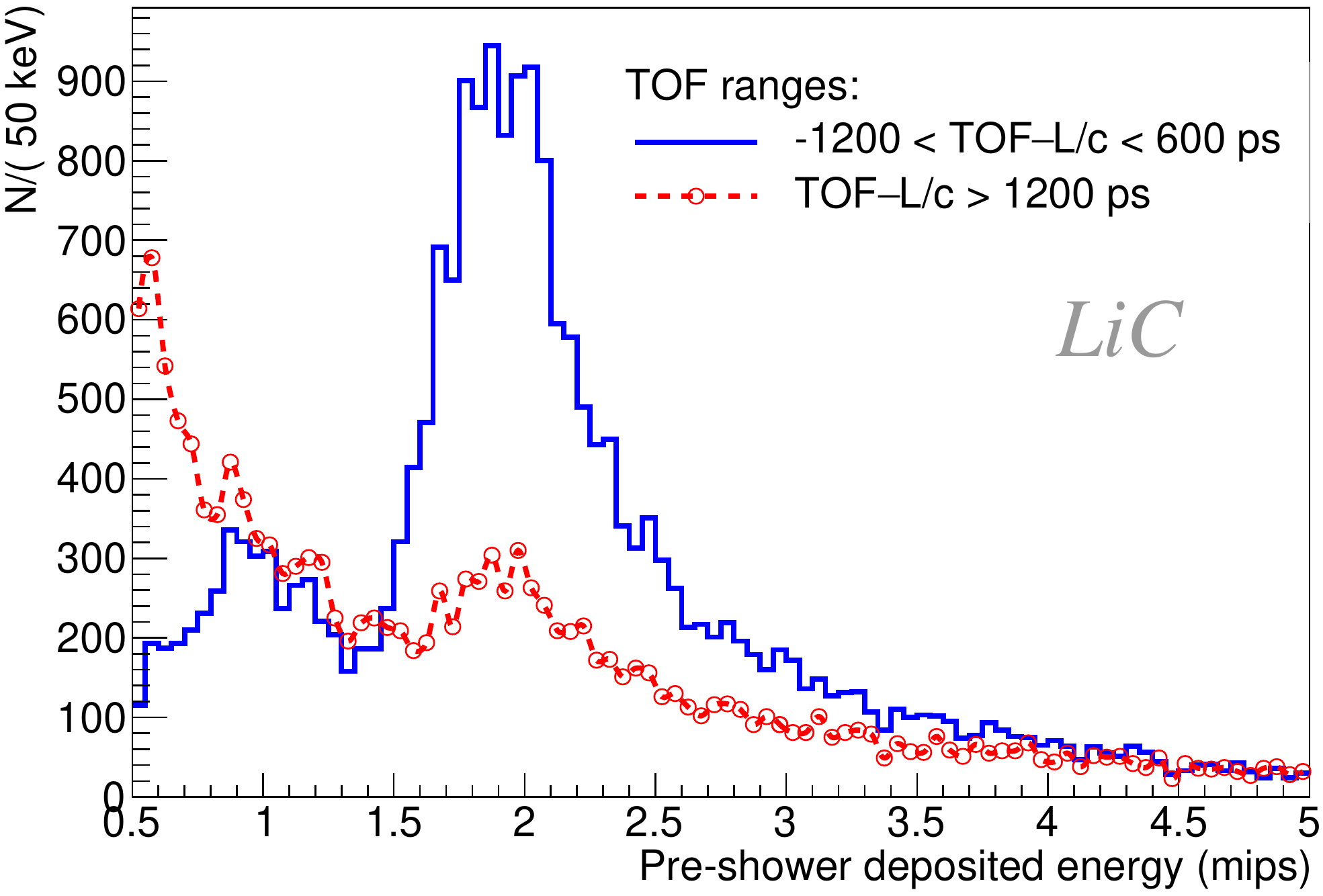}
\caption{Top panel: time of flight of neutral particles
between the beam counter and
the pre-shower for d+C interactions.
Bottom panel: same as above for Li+C interactions.}
\label{fig-4}
\end{figure}

During the Nuclotron run, SPEC is set at an angle 
of 16$^\circ $ relative to the beam direction.
The front
plane of crystals is away from the target at the distance 203 cm. 
The digitization of plastic scintillators
is realised  with a CAMAC ADCs
(Lecroy 2249A) and TDCs (LeCroy 2228A), 
the digitization of analog signals of calorimeter - 
by ADC CC-008.

\begin{figure}
\includegraphics[width=0.5\textwidth]{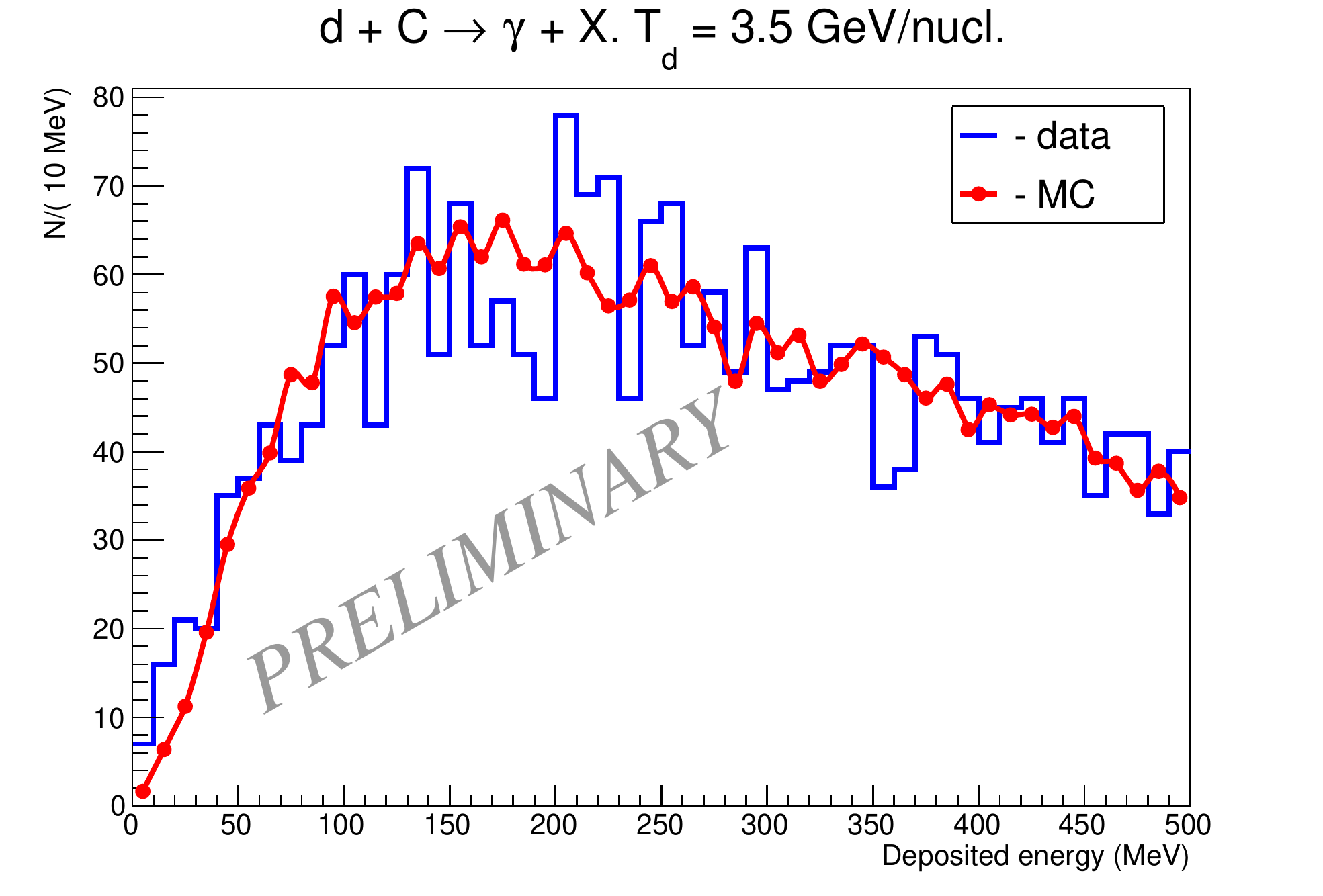}
\includegraphics[width=0.5\textwidth]{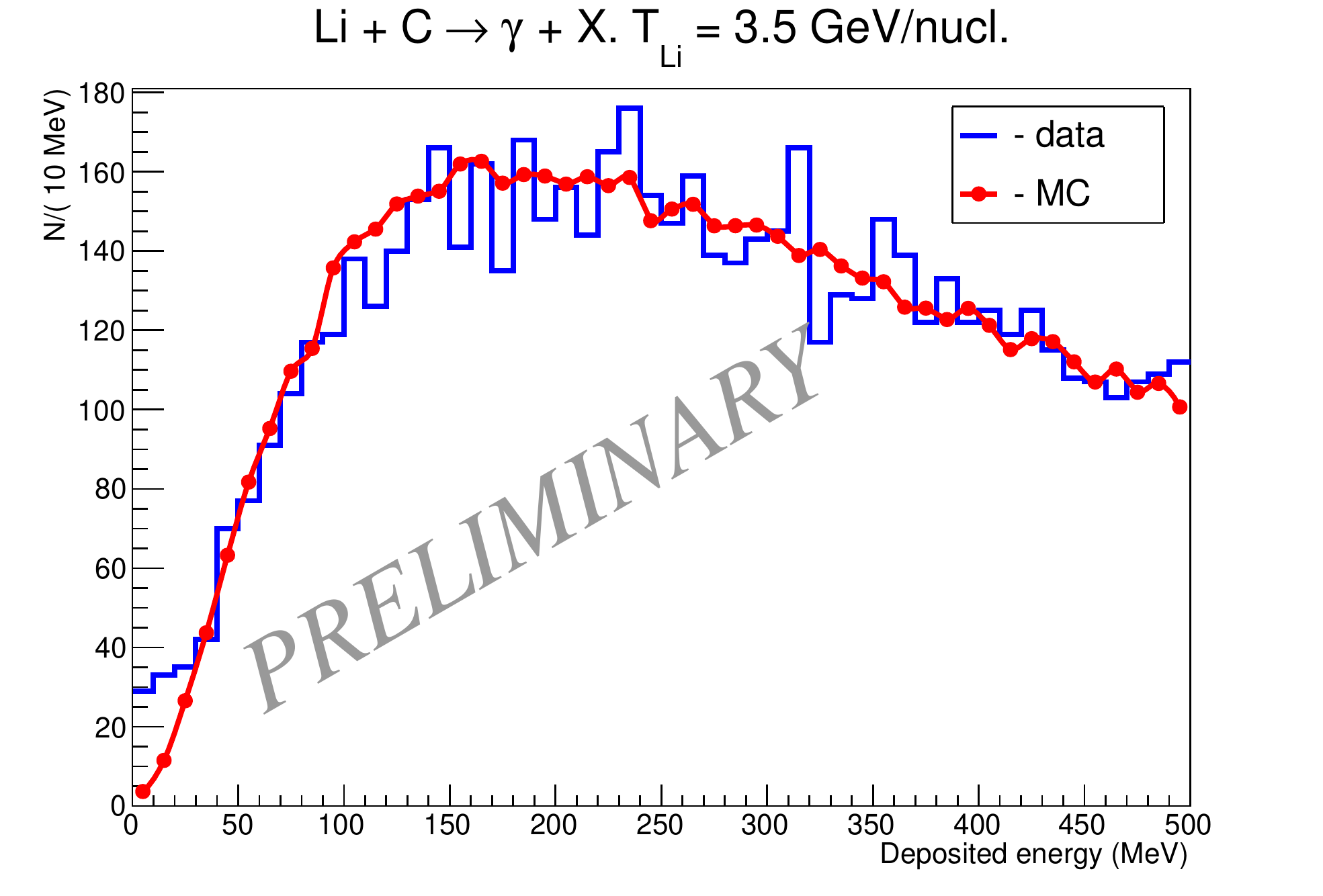}
\caption{Entire energy spectra in SPEC with pre-shower 
and simulation in d+C (a top panel) and
in Li+C (a bottom panel) interactions at Nuclotron.}
\label{fig-5}
\end{figure}

We used CAMAC and a LE-88K crate-controller
with input for a trigger signal. The crate-controller has been
connected to PC with PCI-QBUS interface. 
Data acquisition software has been developed
in MIDAS framework (http://midas.psi.ch).
Time of flight between the beam counter and
the pre-shower for neutral particles 
(no signal in the front-veto) has given
time resolutions 632 ps for d+C and 532 ps for
Li+C interactions (Fig.~\ref{fig-3}).

In 2014 and 2015 two experimental runs (50th and 51st, correspondingly) have been carried out
at Nuclotron in LHEP JINR with 3.5 A GeV deuterium (18 hours) 
and lithium (14 hours) beams. SPEC has been 
installed at the location of NIS-GIBS setup.

Criterions of selection events were as the following:

1) energy in the front veto-counter smaller than 0.3 MIPs;

2) energy in pre-shower 0.5 < $E$ < 4 MIPs;

3) time of flight - 1200 < t - t$_\gamma $ < 600 ps;

4) more than 2 MeV is registered in one of BGO crystals;

5) location of shower in BGO crystal must overlay 
throughout vertical with the triggered pre-shower counter;

6) energy deposition in the outer BGO layer should be no more than 1/3 
of a total to prevent significant leakages. 

In Fig.~4 spectra of $\gamma $ quanta deposited 
in pre-shower plastic with time selection for neutral particles
is presented for d+C (on top) and Li+C (below) interactions. A solid 
line shows Compton peak at 
1~MIP energy and more intensive peak of gamma quanta 
conversion at 2 MIP. In this Fig. with dotted line 
this structure for $t-t_\gamma $ > 1200 ps is almost unnoticeable.

Monte Carlo simulation of the SPEC setup has been 
carried out at the conditions of the last assembly and
the beam energy -- 3.5 A GeV. Monte-Carlo simulation 
(uRQMD+Geant-3.21) is used. Geant-4 has shown
the same result.

After data processing we have obtained SP spectra of
energy release in deuterium-carbon (Fig.~\ref{fig-5}, a top panel) and
lithium-carbon (Fig.~\ref{fig-5}, a bottom panel) interactions. In the region
of energy below than 50 MeV, a noticeable
excess over Monte-Carlo simulation 
has been observed. It agrees well to other SP experiments
\cite{WA83}.

\end{document}